\documentclass[conferences]{IEEEtran}

\usepackage{floatflt}
\usepackage{color}
\usepackage{cite}
\usepackage{graphicx}
\usepackage{amsmath}
\usepackage{amsthm}
\usepackage{amssymb}
\usepackage{verbatim}
\usepackage{psfrag,array}
\usepackage{subfigure}
\usepackage{multirow}
\usepackage{hhline}
\usepackage{stfloats}
\usepackage{amsmath}
\usepackage{epstopdf}
\usepackage{algorithmic}

\newcommand{\p}[1]{\mathop{\mbox{\it p} } }

\renewcommand{\vec}[1]{\ensuremath{\boldsymbol{#1}}}
\newcommand{\be}{\begin{equation}}
\newcommand{\ee}{\end{equation}}
\newcommand{\ba}{\begin{array}}
\newcommand{\ea}{\end{array}}
\newcommand{\bea}{\begin{eqnarray}}
\newcommand{\eea}{\end{eqnarray}}
\newcommand{\bean}{\begin{eqnarray*}}
\newcommand{\eean}{\end{eqnarray*}}

\newcommand{\Ts}{T_\mathrm{s}}
\newcommand{\Fs}{F_\mathrm{s}}

\newcommand{\scell}[2][c]{\begin{tabular}[#1]{@{}c@{}}#2\end{tabular}}

\definecolor{white}{rgb}{1,1,1}

\begin{document}

\title{Improving the Performance of OTDOA based Positioning in NB-IoT Systems}
\author
{
Sha Hu$^{\dagger}$, Axel Berg$^{\ddagger}$, Xuhong Li$^{\dagger}$, and Fredrik Rusek$^{\dagger}$  \\
$^{\dagger}$Department of Electrical and Information Technology, Lund University, Lund, Sweden \\
$^{\ddagger}$ARM, Lund, Sweden \\
$^{\dagger}$\{sha.hu, xuhong.li, fredrik.rusek\}@eit.lth.se, $^{\ddagger}$axel.berg@arm.com
}

\maketitle

\vspace*{-4mm}

\begin{abstract}
In this paper, we consider positioning with observed-time-difference-of-arrival (OTDOA) for a device deployed in long-term-evolution (LTE) based narrow-band Internet-of-things (NB-IoT) systems. We propose an iterative expectation-maximization based successive interference cancellation (EM-SIC) algorithm to jointly consider estimations of residual frequency-offset (FO), fading-channel taps and time-of-arrival (ToA) of the first arrival-path for each of the detected cells. In order to design a low complexity ToA detector and also due to the limits of low-cost analog circuits, we assume an NB-IoT device working at a low-sampling rate such as 1.92 MHz or lower. The proposed EM-SIC algorithm comprises two stages to detect ToA, based on which OTDOA can be calculated. In a first stage, after running the EM-SIC block a predefined number of iterations, a coarse ToA is estimated for each of the detected cells. Then in a second stage, to improve the ToA resolution, a low-pass filter is utilized to interpolate the correlations of time-domain PRS signal evaluated at a low sampling-rate to a high sampling-rate such as 30.72 MHz. To keep low-complexity, only the correlations inside a small search window centered at the coarse ToA estimates are upsampled. Then, the refined ToAs are estimated based on upsampled correlations. If at least three cells are detected, with OTDOA and the locations of detected cell sites, the position of the NB-IoT device can be estimated. We show through numerical simulations that, the proposed EM-SIC based ToA detector is robust against impairments introduced by inter-cell interference, fading-channel and residual FO. Thus significant signal-to-noise (SNR) gains are obtained over traditional ToA detectors that do not consider these impairments when positioning a device. 
\end{abstract}

\section{Introduction}
Observed-time-difference-of-arrival (OTDOA) is a downlink (DL) positioning method first introduced in long-term-evolution (LTE) in Rel. 9 \cite{3GPP, F14}. The positioning-reference-signal (PRS) is transmitted in the DL to enhance positioning measurements at receiver nodes to ensure sufficiently high signal quality and detection probability. The PRS is distributed in time and frequency resources over a subframe and a number of consecutive positioning subframes are allocated with a certain periodicity. In a PRS subframe where PRS is present, no data but only control signaling is transmitted which reduces the interference from neighbour cells \cite{F14}. In order to further reduce inter-cell interference, the network can mute PRS transmission of certain e-NodeBs (termed PRS muting). When PRS is not available, cell-specific-reference-signal (CRS) can also be used to detect the OTDOA. 

\begin{figure}[t]
\vspace{-1mm}
\hspace{20mm}
\scalebox{0.50}{\includegraphics{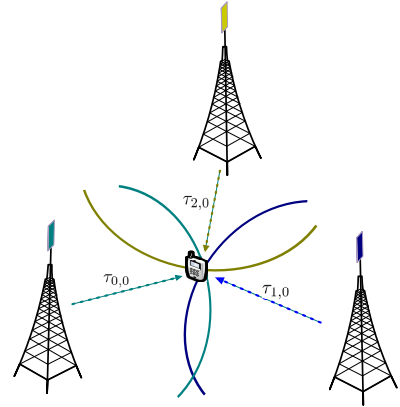}}
\vspace{-2mm}
\caption{\label{fig1}Positioning with PRS signals from 3 e-NodeBs based on OTDOA, where the ToA ($\tau_{0,0}, \tau_{1,0}$ and $\tau_{2,0}$) of the first arrival-paths are detected at an NB-IoT device, and then reported back in the uplink (UL) for e-NodeBs to estimate the position of the device.}
\vspace{-7mm}
\end{figure}

As is well-known, low-cost and power-efficient transceiver circuits are important for a device working in narrow-band Internet-of-things (NB-IoT) systems, which needs to operate for 10 years with its built-in battery. Therefore, using an analog RF circuitry that supports sampling-rate up to 30.72 MHz to obtain good resolution of time-of-arrival (ToA) is expensive and infeasible for a low-end NB-IoT device. Although the device can interpolate the received samples in digital domain to have a high sampling-rate, intensive computations of PRS correlation consume lots of the battery-power. Hence, in this work we consider an NB-IoT device that only works at a low sampling-rate such as 1.92 MHz, and instead of upsampling the received samples we interpolate the PRS correlations in a small search window to improve the ToA resolution.

In order to position an NB-IoT device, PRS signals from at least three cells need to be detected as depicted in Fig. \ref{fig1}. To enhance the hear ability from multiple cells, the positioning subframes are designed with no data transmission in the physical-downlink-shared-channel (PDSCH). However, due to the time-delays, the PRS signal from different cells can still cause strong interference to other cells, which degrades the ToA detection performance. For instance, consider an NB-IoT device that is very close to one e-NodeB; this causes dramatical interference to other cells which results in a positioning failure since only one PRS signal from the closest cell can be detected. Therefore, a successive interference cancellation (SIC) technique is needed for the ToA detection. Although PRS muting can avoid collisions of PRS signal, it increases the latency of positioning process linearly in the number of cells involved and is infeasible for a dense cell-deployment. Besides inter-cell interference, the residual frequency offset (FO) due to an imperfect FO estimate based on NB-IoT synchronization signals also causes performance degradation of the coherent additions of correlations in the ToA detection process. Moreover, although an NB-IoT device is expected to have a low speed and the channel can be assumed constant over one PRS subframe, the wireless fading-channel still needs to be estimated for the SIC process and the coherent additions of correlations.

In this paper, we consider OTDOA based positioning in LTE based NB-IoT systems where only one physical resource block (PRB) is used for data-transmission (180 KHz). We propose an expectation-maximization based successive interference cancellation (EM-SIC) algorithm to estimate the fading-channel, residual FO and ToA of the first arrival-path for each of the detected multiple cells. The EM-SIC algorithm based ToA detector works on received time-domain samples at a low sampling-rate, to firstly obtain coarse estimates of the ToA for all detected cells. Then, the resolution of estimated ToA is refined with upsampled correlations inside a search window centered at the coarse ToA estimates using interpolations. To further improve the accuracies of ToA estimates, we use an iterative multi-path detection (MPD) algorithm to detect the ToA of the first arrival-path by taking into account the property of the auto-correlation function (ACF) of time-domain PRS signal. We show through numerical results that, the proposed EM-SIC algorithm based ToA detector renders significant signal-to-noise (SNR) gains compared to traditional OTDOA detectors that do not thoroughly consider the impairments introduced by the inter-cell interference, ACF of the time-domain PRS signal, fading-channel and residual FO.

\section{PRS and OTDOA based Positioning in NB-IoT Systems}

\subsection{PRS Generation and Subframe Mapping}
\begin{figure}[t]
\begin{center}
\vspace*{-0mm}
\hspace*{-0mm}
\scalebox{1.3}{\includegraphics{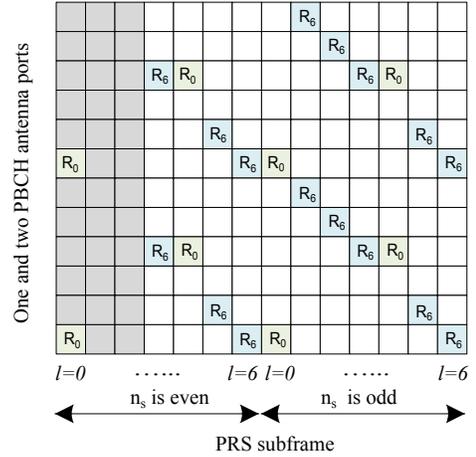}}
\vspace*{-2mm}
\caption{\label{fig2}The PRS pattern in one PRB for normal CP and one or two PBCH antenna ports for a cell with PCI=0 in LTE. The PRS is transmitted on antenna port 6 which is labeled as $R_6$, while the normal CRS is sent on antenna port 0 and labeled as $R_0$. The frequency shift is given by mod (PCI, 6).}
\vspace*{-7mm}
\end{center}
\end{figure}

We consider an LTE based single-input-single-output (SISO) NB-IoT system, where the PRS signal is generated based on physical cell identity (PCI), and mapped to resource element (RE) over a time-frequency grid as described in \cite{3GPP}. The number of consecutive PRS subframes can be either 1, 2, 4 or 6, which are transmitted periodically in the DL. The period of one positioning occasion can be configured to every 160, 320, 640 or 1280 milliseconds (ms). In the considered NB-IoT system, we assume a narrow-band data-transmission over one PRB. The QPSK-modulated PRS signal is generated as \cite{3GPP}
\bea \label{zn} z_{n_s,\ell}=\frac{1}{\sqrt{2}}\left(1-2c[2m]\right)+j\frac{1}{\sqrt{2}}\left(1-2c[2m+1]\right)\!. \eea
where $n_s$ is the slot number within a radio-frame, $\ell$ is the OFDM symbol number within one slot, and $m\!=\! 0, 1$ represents the two PRS symbols in each PRS OFDM symbol. The pseudo-random sequence $c[m]$ is generated by a length-31 Gold sequence whose initial state depends on PCI, $n_s$, $\ell$ and the cyclic-prefix (CP) type. According to the PRS mapping pattern such as shown in Fig. \ref{fig2}, the QPSK symbols are mapped to REs. As there is no data transmission in the PRS subframe, the baseband time-domain PRS signal is obtained through inverse Fast-Fourier-Transform (IFFT) according to
{\setlength\arraycolsep{0pt} \bea  \label{spl} s_{p,\ell}[n]&=&\frac{1}{\sqrt{N}}\!\!\sum_{k=-N/2}^{N/2-1}\!S_{p,\ell}[k]e^{\frac{j2\pi nk}{N}}\notag \\
&=& \frac{1}{\sqrt{N}}\!\left( \!S_{p,\ell}[k_{p,\ell}]e^{\frac{j2\pi nk_{p,\ell}}{N}}\!+\!S_{p,\ell}[k_{p,\ell}\!+\!6]e^{\frac{j2\pi n(k_{p,\ell}\!+\!6)}{N}}\right)\!, \notag \\  \eea}
\hspace{-1.4mm}where $N$ is the IFFT size, and $S_{p,\ell }[k]$ is the mapped PRS signal generated in (\ref{zn}) on the $\ell$th OFDM symbol of the $p$th cell. In total we consider $P$ cells for OTDOA based positioning, and $k_{p,\ell}$ is the lower frequency index where the corresponding PRS signal is transmitted on the $\ell$th symbol of the $p$th cell.

\begin{figure*}[t]
\begin{center}
\vspace*{-10mm}
\hspace*{-6mm}
\scalebox{0.29}{\includegraphics{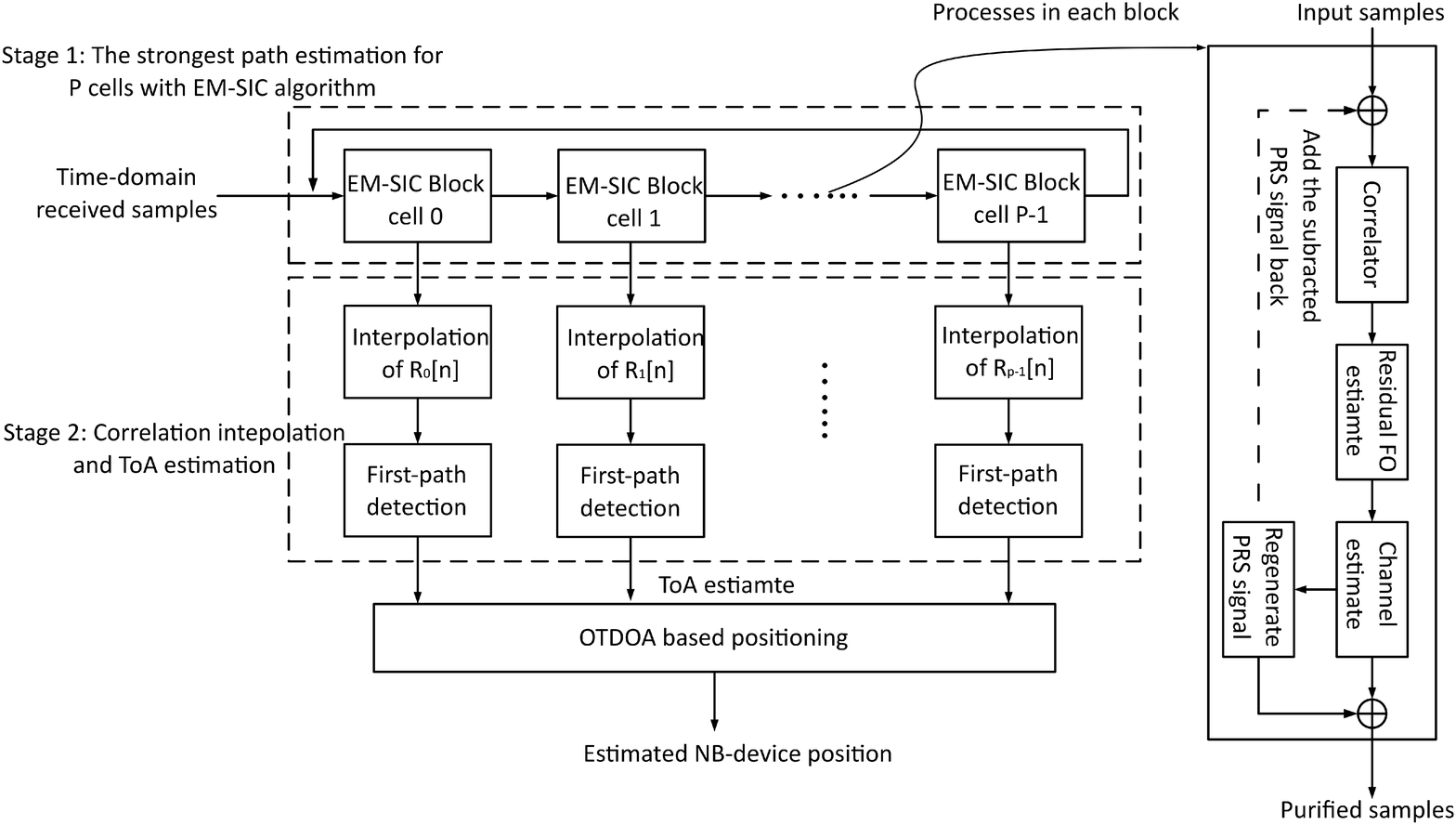}}
\vspace*{-7mm}
\caption{\label{fig3}ToA Detection structure with proposed EM-SIC algorithm.}
\vspace*{-6mm}
\end{center}
\end{figure*}

\subsection{Received Signal Model}
In LTE, the unit of ToA for OTDOA based positioning is $\Ts\!=\!1/\Fs$ where $\Fs\!=\!30.72$ MHz. In the considered NB-IoT system, we assume a sampling-rate $\tilde{F}_{\rm s}$ with a typical value 1.92 MHz, which is much lower than $\Fs$. Denote the true ToA of the $i$th channel-tap from the $p$th cell as $\tau_{p,i}$, and the ToA measured in number of samples as
\bea  n_{p,i}=\lfloor\tau_{p,i}\tilde{F}_{\rm s}\rfloor. \eea
Then, the received signal from the $p$th cell corresponding to the $\ell$th PRS OFDM symbol and $i$th path of the fading-channel at time-epoch $n$ can be modeled as
{\setlength\arraycolsep{0pt} \bea  \label{md1}  y_{p,i}[n\!+\!\ell M]\!=\!\left\{\begin{array}{ll}
0,\;\;&0\!\leq\! n\!<\!   n_{p,i} \\
h_{p,i}^{\ell}[n]s_{p,\ell}\!\left[n\!-\!n_{p,i}\right]\!,\;\;& n_{p,i}\!\leq\! n\!< \! n_{p,i}\!+\!M   \\
0,\;\;&  n_{p,i}\!+M\!\leq\! n\!<\!\tilde{M} \end{array}\right.   \qquad \eea}
\hspace{-1.4mm}where $s_{p,\ell}[n]$ is the time-domain PRS signal (including CP) generated in (\ref{spl}) with $M$ being its length with the sampling-frequency $\tilde{F}_{\rm s}$, and $\tilde{M}$ equals $M$ plus the number of samples corresponding to the maximal ToA for all $P$ cells. The channel $h_{p,i}^{\ell}[n]$ comprises two parts: the $i$th tap of the fading-channel $h_{p,i}$ which we assume to be constant over one PRS subframe, and a phase-rotation caused by the residual FO, which equals
\bea h_{p,i}^{\ell}[n]=h_{p,i}e^{j2\pi \frac{ \epsilon_p(n+\ell M)}{N}}.\eea
where $\epsilon_p$ is the residual FO of the $p$th cell normalized by the subcarrier frequency-spacing (15 KHz). Then, the superimposed received samples for all $P$ cells reads
\bea y[n]=\sum_{p=0}^{P-1}\sum_{\ell=0}^{N_{\mathrm{sym}}}\sum_{i=0}^{L_p-1}y_{p,i}[n\!+\!\ell M]+w[n],\eea
where $w[n]$ is modeled as AWGN with zero-mean and variance $\sigma^2$, and $L_p$ is the maximal number of channel-taps for the $p$th cell. The signal-to-noise (SNR) is defined as $\mathrm{SNR}\!=\!\sigma_{\mathrm{s}}^2/\sigma^2$ where $\sigma_{\mathrm{s}}^2$ is the averaged power of signal $s_{p,\ell}[n]$. We denote the number of OFDM symbols in one subframe as $N_{\mathrm{sym}}$, and the number of OFDM symbols carrying PRS as $N_{\mathrm{PRS}}$. From Fig. \ref{fig2}, $N_{\mathrm{sym}}$ and $N_{\mathrm{PRS}}$ are equal to 14 and 8, respectively. For simplicity, we let $\tilde{\ell}(s)$ ($0\!\leq\!s\!<\!8$) denote the $s$th OFDM symbol that contains PRS. In order to position an NB-IoT device\footnote{Although considering CRS can further improve the positioning performance, in this paper we only consider 8 OFDM symbols on which the PRS are transmitted for positioning. The same principle can be applied to CRS assisted positioning.}, the device needs to detect ToA of the first arrival-path for (at least three out of) the $P$ cells, that is, detect $n_{p,0}$.

\subsection{ToA Detection of the First Arrival-Path}
In order to detect the ToA, for each cell a cross-correlation between the received samples and the PRS signals $s_{p,\ell}[n]$ is implemented, and the correlations are calculated as
\bea \label{Rp} R_{p,\ell}[n]=\!\!\sum_{k=n}^{n+M-1}\!\!y[k]s^{\ast}_{p,\ell}[k-n].\eea
Then, the delay $n_{p,0}$ can be estimated according to
\bea \label{threshold1} n_{p,0}=\underset{n}{\mathrm{argmin\;\;}} \left\{ \frac{\left|\sum\limits_{s=0}^{7}R_{p,\tilde{\ell}(s)}[n]\right|}{\underset{n}{\max} \left\{\left|\sum\limits_{s=0}^{7 }R_{p,\tilde{\ell}(s)}[n]\right|\right\}}>\eta_1 \right\}\!. \eea
However, (\ref{threshold1}) is not executed unless the following is satisfied
\be \label{threshold2} \underset{n}{\max} \left\{\sum\limits_{s=0}^{7}\left|R_{p,\tilde{\ell}(s)}[n]\right|\right\}\!>\!\frac{\eta_2}{\tilde{M}\!-\!M\!-\!1}\!\!\! \!\sum\limits_{n=0}^{\tilde{M}\!-\!M\!-\!1}\!\left|\sum\limits_{s=0}^{7}R_{p,\tilde{\ell}(s)}[n]\right|\!. \ee
For both (\ref{threshold1}) and (\ref{threshold2}), $\eta_1$ and $\eta_2$ are a predefined thresholds that can be adapted for different scenarios. The condition (\ref{threshold2}) decides if the PRS signal is detected while the condition (\ref{threshold1}) provides the estimation of the ToA. A drawback of decision conditions (\ref{threshold1}) and (\ref{threshold2}) is that, the presence of residual FO degrades the performance of coherent addition of $R_{p,\tilde{\ell}(s)}[k]$ over all PRS symbols, and a non-coherent addition renders inferior performance compared to coherent addition.

\subsection{Cram\'{e}r-Rao Lower Bound (CRLB)}
According to (\ref{md1}), under AWGN channel ($h_{p,i}\!=\!0$ for $i\!\neq\!0$) without residual FO, the CRLB for estimating $n_{p,0}$ can be shown to be \cite{XD14}
{\setlength\arraycolsep{2pt} \bea \mathrm{var}(n_{p,0})&\geq& \frac{\sigma^2 N^2}{8\pi^2\sum\limits_{s=0}^{7}\sum\limits_{n=0}^{N-1}\!n^2\left|S_{p,\tilde{\ell}(s)}[n]\right|^2} \notag \\
&=&\frac{\sigma^2 N^2}{8\pi^2\sum\limits_{s=0}^{7}\left(k_{p,\tilde{\ell}(s)}^2+\left(k_{p,\tilde{\ell}(s)}^2\!+\!6\right)^2\right)}, \eea}
\hspace{-1.4mm}which shows that the CRLB for estimating $n_{p,0}$ depends on the subcarrier index $k_{p,\tilde{\ell}(s)}$ where the PRS signal is transmitted. Note that, the value of $k_{p,\tilde{\ell}(s)}$ changes slightly with different FFT shift operations in practical implementations of (\ref{spl}). Although the CRLB is usually difficult to attain unless under the AWGN channel and without impairments caused by interference or FO, it still provides an insight for designing the optimal PRS mapping-pattern of the NB-IoT system.

\begin{figure}[t]
\begin{center}
\vspace*{-3mm}
\hspace*{-8mm}
\scalebox{0.31}{\includegraphics{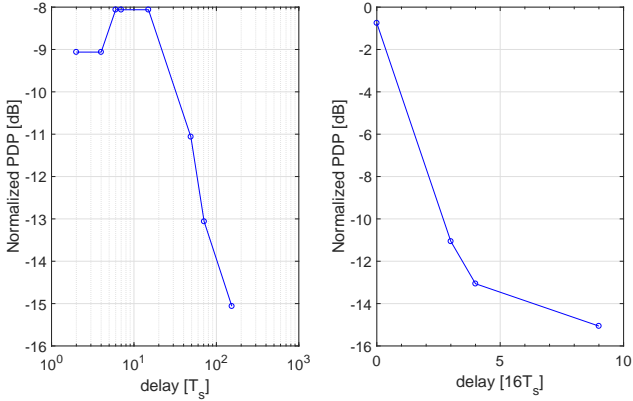}}
\vspace*{-8mm}
\caption{\label{fig4}The normalized power-delay-profile (PDP) of ETU channel with sampling period $\Ts$ and $16\Ts$, respectively.}
\vspace*{-6mm}
\end{center}
\end{figure}

\section{Proposed EM-SIC Algorithm for OTDOA Positioning}
In this section, we elaborate the proposed EM-SIC algorithm based ToA detector, which jointly considers mitigating the inter-cell interference, estimating the residual FO and the multi-path fading-channels, and detecting ToA of all cells.

\subsection{ToA Detection and OTDOA Positioning Structure}
In Fig. \ref{fig3} we depict the detector structure with the EM-SIC algorithm for OTDOA based positioning of the NB-IoT device, which essentially comprises two stages. In the first stage, coarse estimations of ToA of the first arrival-path of all $P$ cells are obtained with iterative EM-SIC blocks, and the correlations are coherently summed-up after the residual FO correction. Then in the second stage, the ToA estimates are refined with interpolated correlations. Note that, as explained earlier, we assume that the EM-SIC algorithm works with the received samples at a low sampling-rate, and as depicted in Fig. \ref{fig4}, the multiple-path components of fading-channel are merged together at a low sampling-rate such as 1.92 MHz. Therefore, for the EM-SIC algorithm in the first stage it is sufficient to only consider the strongest path. 

In the first stage, for each cell an EM-SIC block is applied to firstly cross-correlating the time-domain PRS signal with the received samples to obtain the correlations $R_{p,\tilde{\ell}(s)}[n]$. Then, the residual FO is estimated by cross-correlating the correlations $R_{p,\tilde{\ell}(s)}[n]$ among the 8 PRS symbols in each subframe. After the FOC, the strongest channel-path is found at the delay that maximizes the coherent addition of the correlations, and the corresponding channel-tap is estimated with LMMSE filtering. Finally, with the estimated FO and channel-tap, the time-domain PRS samples are regenerated and removed from the received signal for each of the cells. The same operations are implemented in a successive way until all $P$ cells have been processed. Then in the next EM-SIC iteration, before computing the correlations $R_{p,\ell}[n]$, for each cell the PRS signal subtracted from the previous iteration is added back. Such an EM-SIC based estimation algorithm was first used to decompose the superposed signals for iterative channel estimation and then also applied for PDSCH detection with inter-cell interferences in LTE systems, which was shown to work well \cite{H14}. The EM-SIC in the first stage can be repeated several times to achieve a better performance. As shown in \cite{H14} and observed in numerical simulations, 2-3 iterations can harvest the major gains.

As the sampling-period at 1.92 MHz is $520$ nanoseconds (ns) which corresponds to a maximal resolution of 156 meters (m) with a single cell, the resolutions of ToA estimates are poor in the first stage. Noting that the Fourier transform of a correlation is the power-spectral-density (PSD) which is band-limited to one PRB, to further improve the resolutions, in the second stage the correlations are interpolated to 30.72 MHz (of a resolution 9.75 m). Then, ToA of the first arrival-path is estimated based on upsampled correlations using an iterative MPD approach. After that, the refined ToA estimates are sent to the OTDOA based positioning module which generates an estimate of the position of the NB-IoT device.

\subsection{Stage 1a: Correlation Per OFDM Symbol}
As the operation in the EM-SIC block is the same for each detected cell, we elaborate the operations for the $p$th cell. At a first step, the cross-correlation $R_{p,\ell}[n]$ in (\ref{Rp}) is computed within the maximal delays $\tilde{M}\!-\!M$ for each of the 8 OFDM symbols carrying PRS signal and for all the $P$ cells independently. Before summing $R_{p,\ell}[n]$ over 8 OFDM symbols, we first need to estimate and correct the FO to improve the detection performance.

\subsection{Stage 1b: Residual FO Estimation}
Since there have been FOC operations using the NB-IoT synchronization signals (NPSS and NSSS) at the synchronization steps, the residual FO can be considered relatively small such that the coherent addition inside one OFDM symbols is still applicable. As there are in total 8 OFDM symbols transmitted, the FO $\epsilon_p$ can be estimated based on $R_{p,\tilde{\ell}(s)}[n]$. Note that, as depicted in Fig. \ref{fig2}, the maximal separation between two OFDM symbols carrying PRS is 10 OFDM symbols. Therefore, the largest residual FO, normalized by the subcarrier frequency-spacing, that can be estimated is $\pm0.05$. 

To estimate the residual FO, we cross-correlate $R_{p,\tilde{\ell}(s)}[n]$ at each delay $n$, and the best-linear-unbiased-estimator (BLUE) for $\epsilon_p[n]$ is
\bea \label{rFO} \tilde{\epsilon}_p[n]=\sum_{m=1}^{4}w(m)\phi(m,n) \eea
where $\phi(m,n)$ equals
\be \phi(m,n)=\!\frac{N}{2\pi M(8-m)}\!\!\sum_{s=0}^{7-m}\frac{\arg\!\left\{R_{p,\tilde{\ell}(s)}[n]R^{\ast}_{p,\tilde{\ell}(s\!+\!m)}[n])\right\}}{\tilde{\ell}(s\!+\!m)-\tilde{\ell}(s)},  \ee
and the operation $\arg\{\cdot\}$ returns the angle which belongs to $[-\pi, \pi)$. The combining coefficients $w[m]$ are set to
\bea \vec{w}=[ 0.4762\; \;0.3095\;\; 0.1667\;\;  0.0476], \eea
computed according to \cite[eq. (16)]{MM99}, and achieves the minimum estimation error (MSE) with 8 PRS OFDM symbols. With $\tilde{\epsilon}_p[n]$ estimated in (\ref{rFO}), we implement FOC and coherently add the correlations for 8 PRS OFDM symbols to obtain
\bea  \label{Rnp} R_p[n]=R_{p,\tilde{\ell}(0)}[n]\!+\!\!\sum_{s=1}^{7}\!\!e^{-j2\pi\tilde{\epsilon}_p[n] \frac{M\left(\tilde{\ell}(s)-\tilde{\ell}(0)\right)}{N}}R_{p,\tilde{\ell}(s)}[n]. \eea
Then, the residual FO estimate for the $p$th cell is set to 
\bea \hat{\epsilon}_p=\tilde{\epsilon}_p[\tilde{n}_{p}], \eea
where the index $\tilde{n}_{p}$ is found such that
\bea   R_{p}[\tilde{n}_{p}]=\underset{n}{\max}\left\{\left|R_p[n]\right|\right\}. \eea
Then, the condition (\ref{threshold2}) is verified to decide whether there is a PRS signal received from the $p$th cell. If (\ref{threshold2}) is not satisfied, the detector directly moves to process the next cell and removes the $p$th cell from current EM-SIC iteration.

\subsection{Stage 1c: Channel Estimation and Interference Cancellation}
As we assume that the fading-channel is approximately constant within one subframe, the channel estimate for the strongest channel-path of the $p$th cell is
\bea \hat{h}_{p}=\frac{1}{8M}\sum_{s=0}^{7}\!\!\sum_{k=\tilde{n}_{p}}^{\tilde{n}_{p}+M-1}\!\!\frac{\tilde{y}[k\!+\!\tilde{\ell}(s) M]}{s_{p,\tilde{\ell}(s)}[k\!-\!\tilde{n}_{p}]}e^{-j2\pi \frac{\hat{\epsilon}_p(k+\tilde{\ell}(s) M)}{N}},\eea
where without loss of generality, we assume $s_{p,\ell}[k]\!\neq\!0$ for all $k$. The purified received samples $\tilde{y}[k]$ after removing the regenerated PRS signal from the first $p$ cells equals
{\setlength\arraycolsep{2pt}\bea \label{ty} \tilde{y}[k\!+\!\tilde{\ell}(s) M]&=&y[k\!+\!\tilde{\ell}(s) M] \notag \\ && -\sum_{q=0}^{p-1}\tilde{h}_{q}s_{q,\tilde{\ell}(s)}[k\!-\!\tilde{n}_{q}]e^{-j2\pi \frac{\hat{\epsilon}_q(k+\tilde{\ell}(s) M)}{N}},\qquad \eea}
\hspace{-1.4mm}where the refined channel estimate $\tilde{h}_{p}$ with LMMSE filter is
\bea \tilde{h}_{p}=\frac{\hat{h}_{p}}{1+ \tilde{\sigma}^2_p/|\hat{h}_{p}|^2}, \eea
 and the noise density $\tilde{\sigma}^2_p$ is estimated according to
{\setlength\arraycolsep{2pt} \bea \tilde{\sigma}^2_{p}&=&\frac{1}{8M}\sum_{s=0}^{7}\!\!\sum_{k=\tilde{n}_{p}}^{\tilde{n}_{p}+M-1}\!\Big|\tilde{y}[k\!+\!\tilde{\ell}(s) M]\notag \\
&&\qquad\qquad\quad-\hat{h}_{p}s_{p,\tilde{\ell}(s)}[k-\tilde{n}_{p}]e^{-j2\pi \frac{\hat{\epsilon}_p(k+\tilde{\ell}(s) M)}{N}}\Big|^2.  \qquad \eea}
\hspace{-1.4mm}After all PRS symbols are estimated and removed from $y[k]$, the data $\tilde{y}[k]$ is sent to the EM-SIC block for processing the next cell. As the PRS signal from the $p$th cell has been removed from $\tilde{y}[k]$, the detection performance of the remaining cells is improved, especially under the case that the received signal power from the $p$th cell is stronger than the others. 

The EM-SIC algorithm (comprising the processes in Sec. III B-D) is repeated for the remaining cells until the ToAs have been estimated for all cells. Once done, the received signal $\tilde{y}[k]$ ideally comprises only noise and remaining PRS signals. Hence, at the beginning of the next EM-SIC iteration for each cell, the regenerated and subtracted signal in (\ref{ty}) corresponding to that cell is added back to $\tilde{y}[k]$ for a refined detection as depicted in Fig. \ref{fig3}.

\begin{figure}[t]
\vspace*{-4mm}
\hspace*{-2mm}
\scalebox{0.31}{\includegraphics{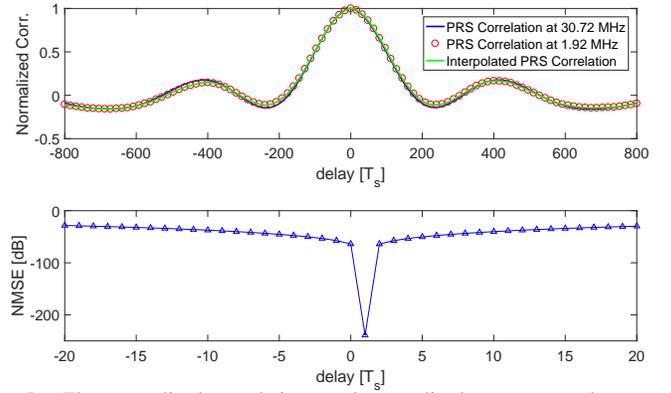}}
\vspace*{-10mm}
\caption{\label{fig6}The normalized correlations and normalized mean-squared-error (NMSE) of the interpolations for one PRS OFDM symbol (without CP) on time-domain under SNR=0 dB at sampling-rate 1.92 and 30.72 MHz, and the interpolated correlations based on 1.92 MHz and with $W\!=\!20$, respectively.}
\vspace*{-6mm}
\end{figure}

\subsection{Stage 2a: Interpolate the Correlations}
As the PSD of the PRS signal is band-limited inside one PRB, we can therefore interpolate the correlations $R_{p}[n]$ to a higher sampling-rate to improve the resolutions of estimated ToAs. With the coarse estimates obtained from the low sampling-rate samples output from the first stage (after SIC and FOC process), the upsampled correlations $\hat{R}_p [m]$ can be interpolated using a \rm{sinc}-function according to
\bea \hat{R}_p [m]=\sum_{n=-W}^{W}R_p [n]\frac{\sin\pi\left(\frac{m}{V}-n\right)}{\pi\left(\frac{m}{V}-n\right)}, \eea
where $V$ is the upsampling rate, and $W$ specifies the window size for searching around the coarse ToA estimate $\tilde{n}_{p}$. As can be seen in Fig. \ref{fig6}, with $R_{p}[n]$ calculated at sampling-rate 1.92 MHz, setting $W\!=\!20$ is sufficient to capture the main-lobe of the normalized PRS correlation function. With setting $V\!=\!16$, the interpolated correlations at 30.72 MHz is also depicted which is shown to be accurate at an SNR of 0 dB.

\vspace*{-4mm}
\subsection{Stage 2b: Iterative MPD}
After obtaining the upsampled correlations $\hat{R}_p [m]$, we can perform the ToA detection of the first arrival-path. Directly comparing the maximal value of $\hat{R}_p [m]$ to a predefined threshold as in (\ref{threshold1}) results in poor performance due to the strong correlations of the PRS signal as shown in Fig. \ref{fig6}. For instance, with an AWGN channel, the ToA estimate should be the index corresponding to the maximal correlation value. However, with a threshold $\eta_1\!<\!1$, the detection (\ref{threshold1}) provides a ToA estimate which is smaller than the true ToA. In order to cope with fading-channels the threshold $\eta_1$ needs to adapt accordingly which is a difficult design task. Instead we utilize a similar method as in \cite{RS16} to implement iteratively MPD of the fading-channel taking into account the ACF of the PRS signal. We claim that a first path is found at position $\tilde{n}_{p,0}$ if
\bea  \label{cond}  \left|\hat{R}_p[\tilde{n}_{p,0}]\right|=\underset{n}{\max}\left\{\left|\hat{R}_p[m]\right|\right\}>\frac{\gamma}{\hat{M}}\sum\limits_{m=0}^{\hat{M}-1} \left|\hat{R}_p[m]\right|, \eea
where $\hat{M}\!=\!V(2W+\!1)$ is the length of $\hat{R}_p[m]$, and $\gamma$ is a predefined peak-to-average (PAR) threshold.

Then we update $\hat{R}_p[m]$ as
\bea  \label{Rnew} \tilde{R}_p[\tilde{n}_{p,0}\!+\!k]=\hat{R}_p[\tilde{n}_{p,0}\!+\!k]-\hat{R}_{p}[\tilde{n}_{p,0}]R_0[k], \eea
where $R_0[k]$ is the normalized ACF of time-domain PRS signal with delay $k$. Then we check again if (\ref{cond}) holds by replacing $\hat{R}_p[m]$ with $\tilde{R}_p[m]$. If so, we claim that a second valid path is present at position $\tilde{n}_{p,1}$ where the maximum of $\left|\tilde{R}_p[n]\right|$ is attained, and then we update $\bar R_p[\tilde{n}_{p,1}\!+\!k]$ again by subtracting the impact of the ACF of the PRS signal according to (\ref{Rnew}). We repeat this process iteratively until the condition (\ref{cond}) is violated, and denote the detected channel-path delays as $\tilde{n}_{p,i}$. Then, the ToA estimate $\tilde{n}_p$ for the $p$th cell is set to
\bea \tilde{n}_p=\underset{i}{\min}\left\{\tilde{n}_{p,i}\right\}\!, \eea
which is the estimated ToA of the detected first arrival-path.

Note that, when path delays of the fading channel are smaller than the main lobe of ACF depicted in Fig. 5, the peak $\tilde{n}_{p,0}$ found in (\ref{cond}) can be an artificial peak caused by overlapping of ACFs corresponding to different channel taps. Therefore, a more accurate ToA estimator shall jointly detect the multi-path components of the fading-channel, which increases the complexity and is out of the scope of this paper.

\begin{figure}[b]
\vspace*{-5mm}
\hspace*{-7mm}
\scalebox{0.3}{\includegraphics{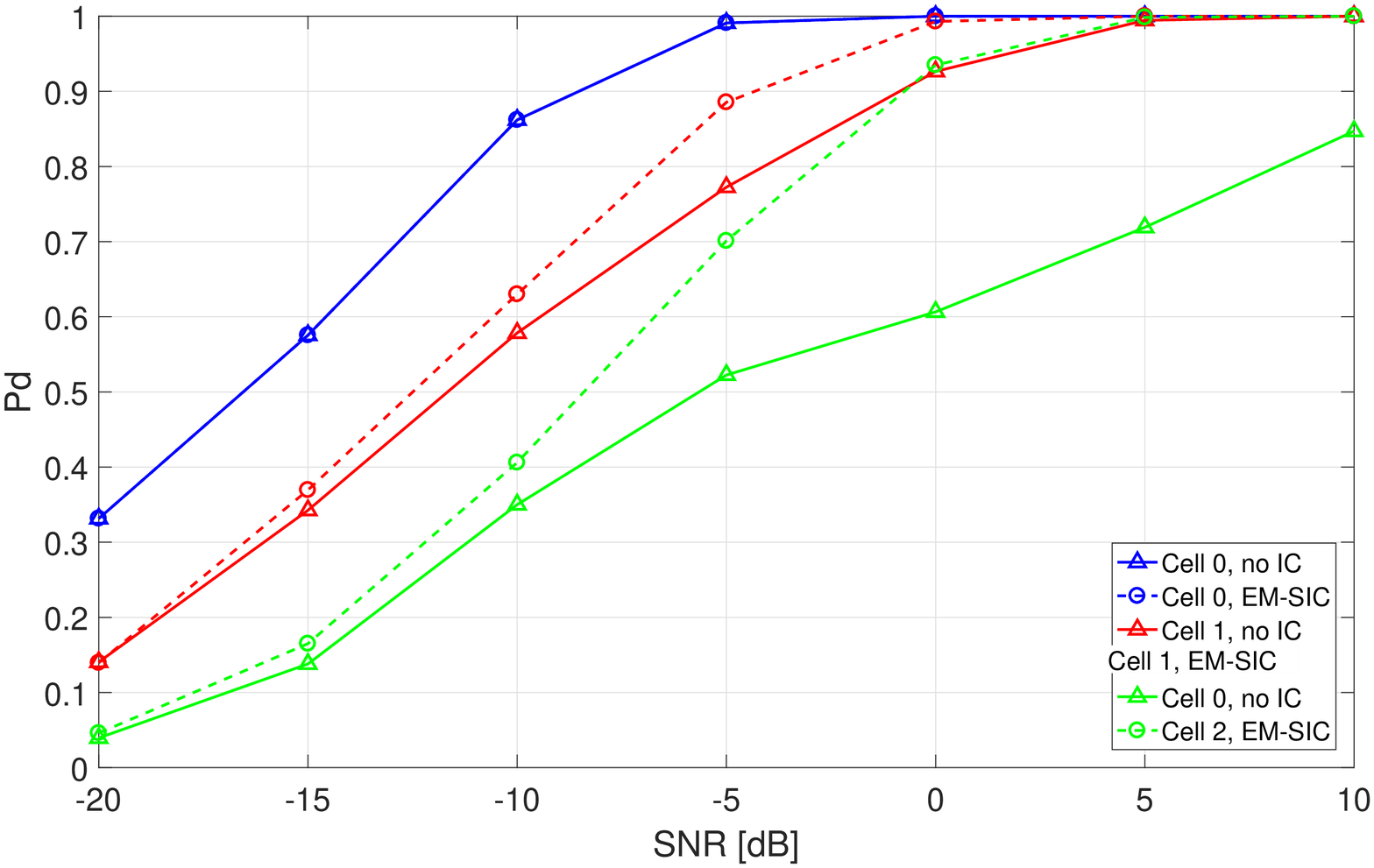}}
\vspace*{-8mm}
\caption{\label{fig7}The detection probability under AWGN channel with 3 cells running the proposed EM-SIC detector only once .}
\vspace*{-3mm}
\end{figure}

\section{Numerical Results}
In this section, we provide numerical results to show the promising performance of proposed EM-SIC based ToA detection for positioning of NB-IoT devices.

\subsection{ToA Detection Performance with 3 Cells}
Firstly we consider ToA detection with 3 cells with PCIs equal to 0, 1 and 2 at sampling rate 1.92 MHz. The ToA $n_{p,0}$ ($0\!\leq\!p\!\leq\!2$) is set to 320, 480 and 640 $\Ts$, and the transmit powers of the three cells are set to 0, -4 and -8 dB, respectively. 

We first evaluate the ToA detection performance without residual FO under AWGN channel, in which case the channel estimation degrades to received power detection. As shown in Fig. \ref{fig7}, compared to a ToA detector without IC, the proposed EM-SIC algorithm provides substantial SNR (measured by the ratio of the transmit power of the strongest cell and the noise density) gains up to 10 dB for the cell that has the least transmit power. 

\begin{figure}
\vspace*{-3mm}
\hspace*{-3mm}
\scalebox{0.3}{\includegraphics{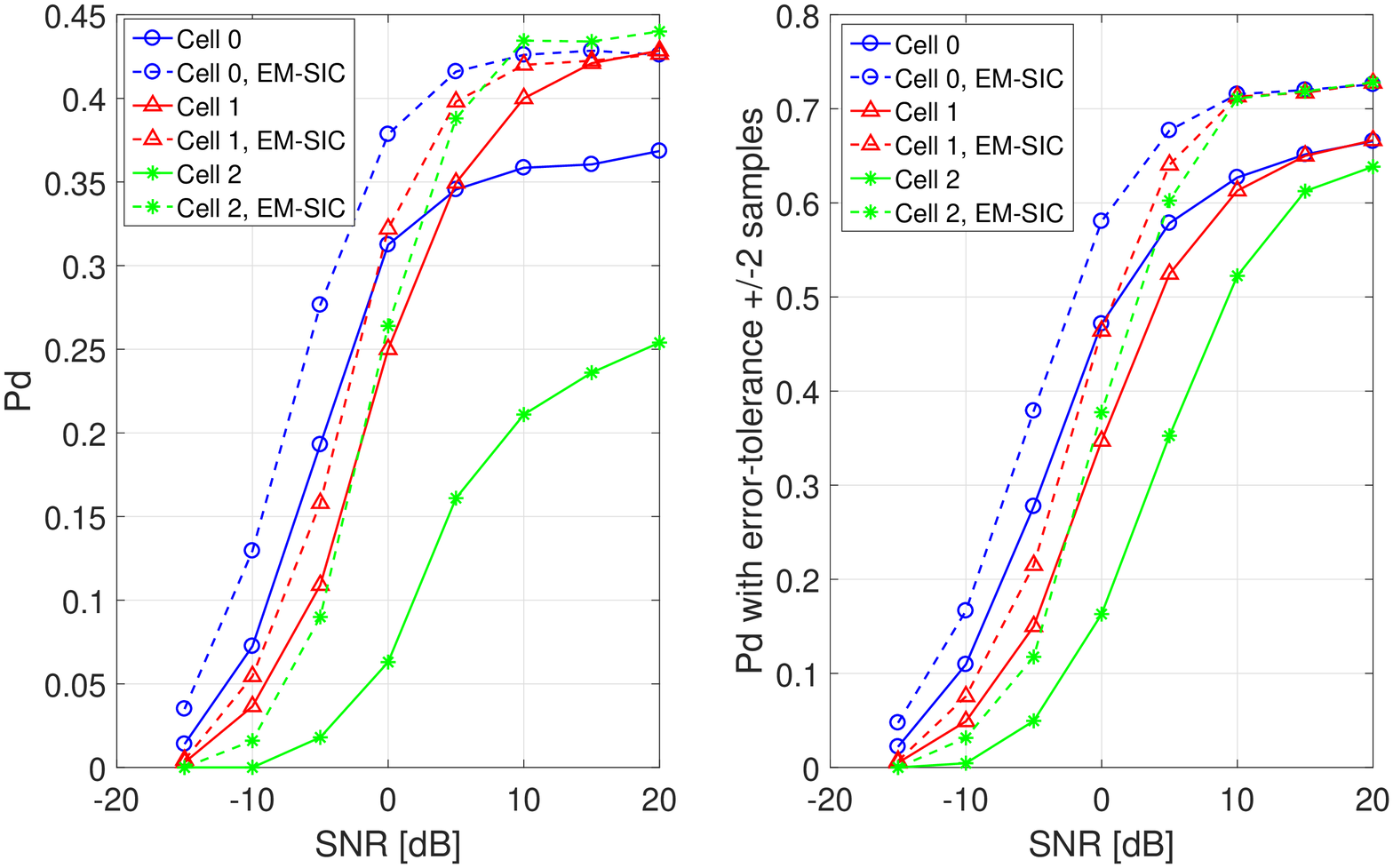}}
\vspace*{-8mm}
\caption{\label{fig8}The detection probability under ETU-3Hz channel for 3 cells with the proposed EM-SIC detector with 2 global iterations.}
\vspace*{-5mm}
\end{figure}

In Fig. \ref{fig8} we repeat the tests under ETU-3Hz channel, and set the normalized residual FO for the three cells to 0.02, 0.01 and 0.01, respectively. As can be seen, the proposed EM-SIC detector renders significant detection improvements compared to the detector that does not apply FOC or SIC processes.

\subsection{ToA Detection Performance in Practical Scenarios}
\begin{figure}
\vspace*{-5mm}
\vspace*{-0mm}
\hspace*{-9mm}
\scalebox{0.33}{\includegraphics{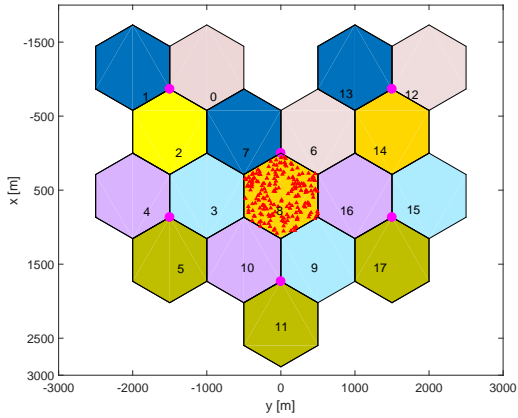}}
\vspace*{-9mm}
\caption{\label{fig9}The geometric deployment of the 6 e-NodeBs (marked as magenta circles and each of them comprises 3 sectors) and 200 NB-devices (marked as red triangles) that are associated to the cell with PCI 8.}
\vspace*{-2mm}
\end{figure}

Next we consider an NB-IoT system with simulation parameters listed in Table I, and the OTDOA based positioning performance is evaluated with 6 cells with a deployment geometry depicted in Fig. \ref{fig9}. We uniformly generate 200 NB-IoT devices associated to the cell with PCI 8. In all simulations we set the thresholds $\eta_2\!=\!0.8$, $\eta_2\!=\!3$ and $\gamma\!=\!7$ and make no efforts to further optimize them for each individual channel condition. At the beginning of all detecting methods we use (\ref{threshold2}) to decide whether PRS signal is present or not, and the false alarm probability of detecting the PRS are similar for all evaluated detectors.

\begin{figure} 
\vspace*{-3mm}
\vspace*{-0mm}
\hspace*{-8mm}
\scalebox{0.31}{\includegraphics{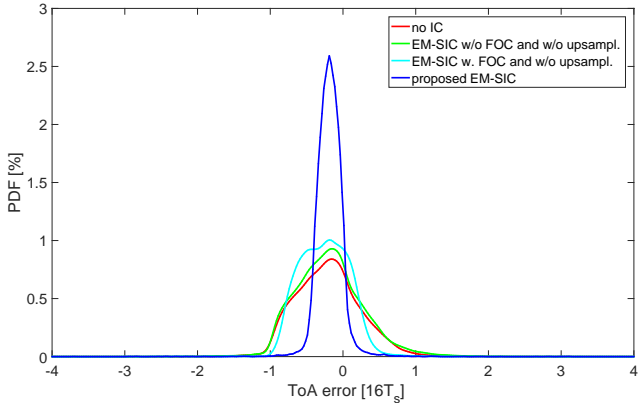}}
\vspace*{-8mm}
\caption{\label{fig10}The PDF of ToA estimation errors. The iterative MPD performs better than the threshold based detection (\ref{threshold1}) under AWGN channel.}
\vspace*{-5mm}
\end{figure}

In Fig. \ref{fig10} we evaluate the probability density function (PDF) of the ToA estimation error when only transmitting the PRS from the cell with PCI 8 and disabling the other cells under AWGN channel. The estimation errors are calculated by subtracting the true ToA with estimated ToA values. As can be seen, the FOC and upsampling using interpolations improves the accuracy of the ToA estimation significantly.
\begin{table}[b]
\renewcommand{\arraystretch}{1.2}
\vspace*{-2mm}
\centering
\caption{Simulation Parameters for OTDOA-based Positioning .}
\label{tab1}
\hspace*{1mm}
\vspace*{-2mm}
\begin{tabular}{|c||c|}
 \hline
Parameter & Value \\ \hhline{|=||=|} 
Number of e-NodeBs& 6  \\ \hline
Inter-cite distance&1.732 km \\ \hline
Frequency band &900 MHz \\ \hline
Channel model & AWGN, ETU   \\ \hline
Number of e-NodeB antenna & 1  \\ \hline
Number of device antenna& 1  \\ \hline
Macro transmit power& 46 dBm for 1.92 MHz   \\ \hline
Thermal noise density& \scell{-174 dBm/Hz for AWGN \\ -184 dBm/Hz for ETU} \\ \hline
Path loss model \cite{R1} ($d$ in km) & $L\!=\!120.9\!+\!37.6\log_{10}(d)$   \\ \hline
Shadowing standard deviation & 8 dB  \\ \hline
Shadowing correlation &\scell{Between e-NodeBs: 0.5 \\
Between sectors of e-NodeB: 1.0} \\ \hline
\scell{Number of PRB / PRS occasion\\ / consecutive PRS subframes}&1\,/1\,/1   \\ \hline
PRS Muting & False   \\ \hline
Normalized residual FO&Uniformly drawn in [-0.03, 0.03].   \\ \hline
\end{tabular}
\vspace{-1mm}
\end{table}

\subsection{OTDOA based Positioning Performance}

\begin{figure}[t]
\vspace*{-4mm}
\vspace*{-0mm}
\hspace*{-3mm}
\scalebox{0.3}{\includegraphics{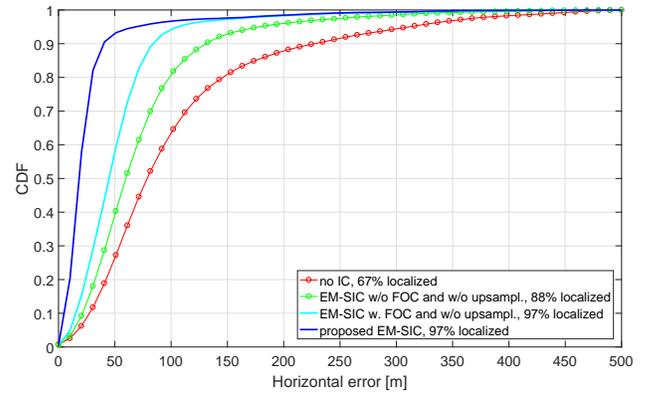}}
\vspace*{-8mm}
\caption{\label{fig11}The OTDOA based positioning performance and localization ratio under AWGN channel.}
\vspace*{-2mm}
\end{figure}

\begin{figure}
\vspace*{-2.5mm}
\vspace*{-0mm}
\hspace*{-3mm}
\scalebox{0.3}{\includegraphics{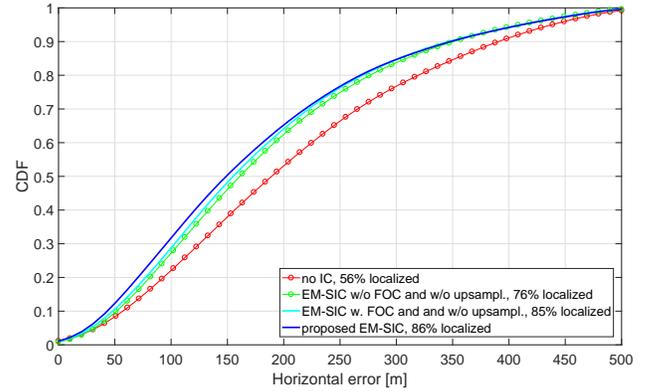}}
\vspace*{-8mm}
\caption{\label{fig12}The OTDOA based positioning performance and localization ratio under ETU-3Hz channel.}
\vspace*{-5mm}
\end{figure}

In Fig. \ref{fig11} and Fig. \ref{fig12} we plot the cumulative distribution function (CDF) of the horizontal positioning error under AWGN channel and ETU-3Hz channels, respectively. The localization percentages of different detection methods are also shown in the legends, where we claim a successful positioning only when at least three cells are detected and the positioning error-distance is less than 500 m. As can be seen, under both channels the proposed EM-SIC detector performs much better than a traditional no IC detector that does not consider the impairments of the interference, residual FO and fading-channels. Moreover, as can be seen, the FOC and upsampling processes greatly improve the positioning performance under AWGN channels, while under EUT-3Hz channel, the gains in positioning accuracy introduced by FOC and upsampling processes are marginal due to inaccurate ToA estimates. Nevertheless, with FOC process the localized ratio is still greatly improved, while the upsampling process further slightly improves that.

\section{Summary}
We have considered the observed-time-difference-of-arrival (OTDOA) based positioning for an NB-IoT device with multiple cells. We have proposed an expectation-maximization based successive interference cancellation (EM-SIC) detector to jointly consider the fading-channel, residual frequency offset (FO), and time-of-arrival (ToA) of the first arrival-path. To design a low complexity ToA detector, the EM-SIC algorithm works with received time-domain samples at a low sampling-rate. The resolution of estimated ToA is further refined by upsampling the correlations obtained at the low sampling-rate inside a small search window to a higher sampling-rate with interpolation. Numerical simulations have shown that, the proposed EM-SIC ToA detector performs robustly against impairments introduced by inter-cell interference, fading-channel and residual FO, which shows significant signal-to-noise (SNR) gains compared to traditional ToA detectors that do not thoroughly consider these impairments.


\begin{thebibliography}{99}
\bibitem{F14} S. Fischer, ``Observed time difference of arrival (OTDOA) positioning in 3GPP LTE,\rq\rq{} White Paper, Qualcomm Technologies Inc., Jun. 2014.

\bibitem{3GPP} 3GPP TS 36.211, ``Evolved Universal Terrestrial Radio Access (E-UTRA): Physical channels and modulation,\rq\rq{} Release 14, Dec. 2016.

\bibitem{LS14} J. Liu  and S. Feng, ``RSTD performance for small bandwidth of OTDOA positioning in 3GPP LTE,"  IEEE Veh. Tech. Conf. (VTC-Fall), Las Vegas, Sep. 2013, pp. 1-5.

\bibitem{XD14} W. Xu, M. Huang, C. Zhu, and A. Dammann, ``Maximum likelihood TOA and OTDOA estimation with first arriving path detection for 3GPP LTE system,"  \textit{Trans. Emerging Tel. Tech.,} vol. 27, no. 3. pp.  339-356, Mar. 2016.


\bibitem{K93} S. M. Kay, \lq\lq{}Fundamentals of statistical signal processing, volume {I}: Estimation theory,\rq\rq{} Prentice Hall signal processing series, 1993.

\bibitem{RS16} H. Ryd\'{e}n, A. A. Zaidi, S. M. Razavi, F. Gunnarsson, and I. Siomin, ``Enhanced time of arrival estimation and quantization for positioning in LTE networks," IEEE Int. Symp. on Personal, Indoor and Mobile Radio Commun. (PIMRC), Valencia, Sep. 2016, pp. 1-6.

\bibitem{H14} S. Hu, G. Wu, B. Priyanto, F. Rusek, S. Kant, and J. Chen, \lq\lq{}Iterative interference cancellation method\rq\rq{}, Patent US20140369300A1, filed on Aug. 2014.

\bibitem{MM99} M. Morelli and U. Mengali, \lq\lq{}An improved frequency offset estimator for OFDM applications,\rq\rq{} \textit{IEEE Comm. Lett.,} vol. 3, no. 3, pp. 75-77, Mar. 1999.

\bibitem{R1} 3GPP TSG RAN WG1, R1-168310, \lq\lq{}WF on simulation assumption for NB-IoT positioning,\rq\rq{} Gothenburg, Sweden, Aug. 22-27, 2016.


\end{thebibliography}
\end{document}